\RequirePackage{fix-cm}
\documentclass[referee,sn-basic,Numbered]{sn-jnl}
\usepackage{etoolbox}
\usepackage{lineno,hyperref}

\let\oldbibliography\thebibliography
\renewcommand{\thebibliography}[1]{\oldbibliography{#1}\setlength{\itemsep}{0pt}}

\modulolinenumbers[5]

\usepackage{flushend}
\usepackage[utf8]{inputenc}
\usepackage{todonotes}
\newcounter{todocounter}

\usepackage{color,soul}
\definecolor{lightgray}{rgb}{0.83, 0.83, 0.83}

\makeatletter
\newcommand\rewblue[1]{\textcolor{black}{#1}}
\def\rewred2#1{\textcolor{red}{#1}}
\DeclareRobustCommand\rew{\@ifnextchar2{\rewred}{\rewblue}}
\makeatother

\newtheorem{theorem}{Theorem}
\newtheorem{proposition}{Proposition}

\usepackage{amsfonts}
\usepackage{url}
\usepackage{caption}
\usepackage{subcaption}

\usepackage{lmodern}

\usepackage{amsmath}
\usepackage{xspace}
\usepackage{multirow}
\usepackage{amssymb}
\usepackage{algorithm}
\usepackage[noend]{algpseudocode}
\usepackage{booktabs}
\renewcommand{\algorithmicrequire}{\textbf{Input:~}}
\renewcommand{\algorithmicensure}{\textbf{Output:~}}

\usepackage{array}
\usepackage{color}
\usepackage{listings}
\usepackage{placeins}
\usepackage{enumitem}
\usepackage{mathtools}
\usepackage{extarrows}
\usepackage{mathrsfs}
\algnewcommand{\algorithmicforeach}{\textbf{for each}}
\algdef{SE}[FOR]{ForEach}{EndForEach}[1]
  {\algorithmicforeach\ #1\ \algorithmicdo}{\algorithmicend\ \algorithmicforeach}

\usepackage{subfiles}

\let\cline\cmidrule
\begin{document}

\title{Accelerating Spanning-Centrality Approximation with Sampling-Aware Vectorized Random Walks}

\author[1]{\fnm{G\"{o}khan} \sur{ G\"{o}kt\"{u}rk}}\email{ggokturk@sabanciuniv.edu}

\author*[1]{\fnm{Kamer} \sur{Kaya}}\email{kaya@sabanciuniv.edu}

\affil[1]{\orgdiv{Faculty of Engineering and Natural Sciences}, \orgname{Sabanc{\i} University}, \orgaddress{\street{Orta Mah.}, \city{Istanbul}, \postcode{34956}, \state{Tuzla}, \country{Türkiye}}}

\abstract{
Spanning centrality quantifies an edge's importance through its share of a graph's spanning trees, and approximating it for all edges~({AESC}) requires processing large volumes of irregular random walks under a prescribed error bound. \rew{Its applications include phylogenetic analysis and network robustness assessment.} We focus on this random-walk processing bottleneck and propose {\em Sampling-Aware Bouquet Arrangement}~(SABA), a \rew{CPU sampling and scheduling scheme} that orders \rew{walks initialized on a uniform grid} so that adjacent walks execute together in SIMD-width groups, called bouquets; \rew{for fixed per-walk random inputs, the ordering preserves the multiset of sampled walks while improving locality during neighbor access.} \rew{On synthetic benchmarks, SABA achieves geometric-mean speedups of $6.95\times$ and $7.41\times$ over traditional random-walk generation for $2048$ and $16{,}384$ walks per vertex, respectively, and up to $\approx49\times$ in the latter setting.} \rew{Within AESC, the complete 16-thread SABA implementation is $58.23\times$ faster than the original single-threaded {\tt TGT+} implementation on average.}
}

\maketitle

\section{Introduction}

{\em Spanning centrality} ({\sc SC}) measures the importance of an edge in a graph for maintaining the graph's connectivity. For an edge $e$ in a connected graph $G$, it is defined as the fraction of spanning trees of
$G$ that contain $e$, i.e.,
$$\textsc{SC}(e) = \frac{\text{Number of spanning trees in $G$ containing } e}{\text{Total number of spanning trees in } G}.$$
\noindent The metric quantifies how crucial an edge is for the graph to remain connected, reflecting the edge's role in the overall structure and stability of the network. \rew{SC has been used in phylogenetic analysis~\cite{Teixeira2016} and to identify edges whose removal disrupts information propagation~\cite{mavroforakis2015spanning}.}

Local metrics such as {\em degree centrality} and distance-based metrics such as {\em closeness} and {\em betweenness centrality} exist in the literature.
Although spanning centrality appears to be {\em yet another metric}, it is global and its focus is beyond the shortest paths. Hence, it is useful to analyze the importance and/or redundancy of all the edges for applications such as {\em phylogeny analysis} or {\em resiliency/robustness analysis}~\cite{Teixeira2016}. \rew{This gives a different view of edge importance from shortest-path-based metrics.}

Let $G = (V, E)$ be an undirected graph with $|V| = n$ vertices and $|E| = m$ edges. As expected, computing {\sc SC}  by taking all the spanning trees into account is not computationally feasible. In the literature, the fastest algorithm to compute the spanning centrality of all the edges in a graph has time complexity $\mathcal{O}(mn^{\frac{3}{2}})$, which is not practical for large networks. This is why {\em approximation algorithms} have been the standard practical approach, and various algorithms have been proposed in the last decade~\cite{aesckdd,peng2021local,hayashi2016efficient,mavroforakis2015spanning}.
\rew{Existing approaches include uniform spanning-tree sampling~\cite{hayashi2016efficient}, random projections with Laplacian solvers~\cite{mavroforakis2015spanning}, and truncated random walks~\cite{peng2021local}. {\tt TGT+}~\cite{aesckdd} combines truncated deterministic traversals with random-walk sampling while keeping every visited vertex available to the application; this property makes it the integration target of our walk-acceleration scheme.}

 As the exact {\sc SC} algorithms suffer from computationally expensive matrix kernels, several approximation algorithms also incur the cost of processing many long random walks. In graphs, a {\em random walk} is a path that traverses the vertices of a graph {\em randomly}~\cite{pearson1905problem}.
 They have been proven to be extremely useful in many randomized algorithms.
   In physics, they are used to model the behaviour of particles or molecules in a medium. In network science, random walks are used to detect communities and simulate the spread of information or diseases. This work focuses on their role in approximating spanning centrality.

 Recently, Zhang~et~al. proposed two approximation algorithms, {\tt TGT} and {\tt TGT+}, to efficiently approximate the spanning centrality of edges in large graphs
~\cite{aesckdd}.
A major contribution is the improved theoretical bounds for truncating long random walks, which significantly enhances the efficiency and accuracy of the approximation process. By optimizing truncated lengths and reducing the number of random paths processed, {\tt TGT+} achieves substantial performance improvements over the state-of-the-art. \rew{Even so, these costs remain prohibitive in practice, as our own measurements on large graphs confirm (Section~\ref{sec:aesc-exp}).}

\rew{Based on the above, three research challenges are addressed in this study: First, since each step of a walk samples a uniformly chosen neighbor, the memory accesses incurred when many walks are processed are irregular, and the memory subsystem dominates the runtime. Second, the outcome of each random selection is required before the next vertex can be visited, so its latency cannot be hidden. Third, the traditional scheme lengthens a single walk at a time, which precludes vectorized processing on modern CPUs.}

In this work, we focus on how to approximate the spanning centrality of the edges in large networks by processing random walks much faster on modern CPUs with SIMD instructions, wide registers, and vectorization opportunities. The contributions of this study are summarized below:

\begin{itemize}
\item A novel hash-based sampling that improves data locality by clustering random walks into bouquets is proposed.
\item A vectorized random-walk processing scheme based on this sampling is designed and implemented.
\item The proposed techniques can be applied to various path-based Monte Carlo algorithms on graphs. \rew{On the synthetic benchmarks, SABA achieves geometric-mean speedups of $6.95\times$/$7.41\times$ over naive random-walk generation for $2048$/$16{,}384$ walks per vertex, and up to $\approx49\times$ in the latter setting.}
\item{\rew{Within the AESC pipeline, SABA provides a $1.50\times$ single-thread speedup over {\tt Baseline}. The complete 16-thread SABA implementation is $58.23\times$ faster than the original single-threaded {\tt TGT+} implementation and $13.42\times$ faster than single-threaded {\tt Baseline}.}}
\item{\rew{The code is publicly available at \url{https://github.com/ggokturk/tgtp}.}}
\end{itemize}

This paper is organized as follows: We define the notation and provide the information on spanning centrality in Section~\ref{sec:back}. Section~\ref{sec:method} presents our proposed method. The implementation details and performance characteristics are explained in Section~\ref{sec:impdetail}.
Sections~\ref{sec:exps} and~\ref{sec:rel} present the experimental results and related work, respectively. Finally, Section~\ref{sec:conc} concludes the paper.

\section{Background and Notation}\label{sec:back}

\begin{figure}[!h]
    \centering
    \includegraphics[width=0.25\linewidth]{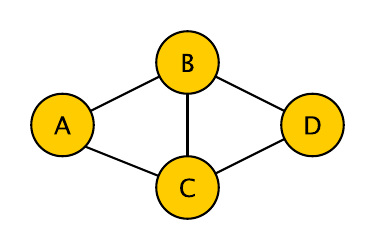}
    \caption{A sample graph with four vertices and five undirected edges. Vertices $A$ and $D$ are degree-2 vertices, and $B$ and $C$ are degree-3 vertices; the graph has eight spanning trees. Edge $\{B,C\}$ is contained in four of these spanning trees, so $\textsc{SC}(\{B,C\})=\frac{1}{2}$, whereas each of the other edges~(e.g., $\{A,B\}$) is contained in five, i.e., $\textsc{SC}=\frac{5}{8}$.}\label{fig:toy}
\end{figure}

Let $G = (V, E)$ be an undirected graph where {\em vertex} set $V$ has $|V|=n$ {\em vertices} and {\em edge} set $E$ has $|E|=m$ {\em edges}. $\Gamma(v) = \{u: \{u,v\} \in E\}$ denotes the neighborhood of $v$ and the degree of $v$ is denoted as $d(v) = |\Gamma(v)|$. A sample graph with 4 vertices and 5 undirected edges is given in Figure~\ref{fig:toy}. The important notations we use in the paper are summarized in Table~\ref{tab:notation}.

  \begin{table}[!ht]
  \renewcommand{\baselinestretch}{1.2}
        \centering
        \caption{Notation used in the paper}
        \label{tab:notation}
        \begin{tabular}{|l|p{0.63\linewidth}|}
            \hline
            Variable & Definition  \\
            \hline
            $G = (V,E)$         & Graph $G$ with vertices $V$ and edges $E$ \\
            $\Gamma(v)$    & Neighborhood of the vertex $v$ in $G$ \\
            $d(v)$              & Degree of the vertex $v$, i.e., $|\Gamma(v)|$ \\
             $\textsc{SC}(e)$ & The spanning centrality of $e $ \\
             $T(G)$ & The set of spanning trees of $G$ \\

$\epsilon$& The absolute error threshold\\
$\delta$& The failure probability\\
$\omega$& The number of eigenvectors used\\
$\gamma$& \rew{The number of candidate vertices} \\
              {$p_l(u,v)$} & \rew{Probability that a walk from $u$ is at $v$ after $l$ steps} \\
             $B$ & Number of random walks per bouquet, i.e., SIMD vector width \\
    $\mathcal{X}_W$ & \rew{Evolving 32-bit state of walk $W$, initialized on a uniform grid} \\
    $h_{max}$ & Hash-domain size, $2^{32}$; values range from $0$ to $h_{max}-1$ \\
    $\beta$ & Size of the pool candidates are drawn from at a bouquet level \\
    $\eta$ & Distinct items within one level of a bouquet \\
            \hline
        \end{tabular}
    \end{table}

\subsection{Spanning Centrality}

\rew{Spanning centrality measures edge importance through graph-wide connectivity. Its applications include phylogenetic analysis~\cite{Teixeira2016} and studies of network resilience and information propagation~\cite{mavroforakis2015spanning}.}

For a connected undirected graph $G = (V, E)$, the {\em spanning centrality} of an edge $e \in E$, $\textsc{SC}(e)$ is defined as the fraction of spanning trees of $G$ that contain $e$ relative to the total number of spanning trees of the entire graph $G$. Hence, when $\textsc{SC}(e) = 1$, the subgraph obtained by removing $e$ from the edge set, i.e., $G \setminus \{e\}$, is disconnected. Formally,
\[\textsc{SC}(e) = 1 - \frac{|T(G \setminus \{e\})|}{|T(G)|} \]
\rew{where $T(G)$ denotes the set of spanning trees of $G$ and $|T(G)|$ their number; this is equivalent to the ratio form used in the introduction, $\textsc{SC}(e) = |\{T \in T(G): e \in T\}|/|T(G)|$. In particular, $\textsc{SC}(e)=1$ precisely when $e$ is a bridge.} \rew{For disconnected inputs, the definition is applied within each edge's component.} Unfortunately, $T$, therefore \textsc{SC}, is expensive to compute, especially for real-world networks.

\rew{As a worked example, consider the sample graph in Figure~\ref{fig:toy}, which has $|T(G)|=8$ spanning trees. Four of them contain $\{B,C\}$, namely $\{\{A,C\},\{B,C\},\{B,D\}\}$, $\{\{A,B\},\{B,C\},\{B,D\}\}$, $\{\{A,C\},\{B,C\},\{C,D\}\}$, and $\{\{A,B\},\{B,C\},\{C,D\}\}$; hence $\textsc{SC}(\{B,C\}) = 1-\frac{4}{8} = \frac{1}{2}$, i.e., removing $\{B,C\}$ leaves the remaining four trees. Five spanning trees contain $\{A,B\}$: $\{\{A,B\},\{A,C\},\{B,D\}\}$, $\{\{A,B\},\{A,C\},\{C,D\}\}$, $\{\{A,B\},\{B,C\},\{B,D\}\}$, $\{\{A,B\},\{B,C\},\{C,D\}\}$, and $\{\{A,B\},\{B,D\},\{C,D\}\}$; hence $\textsc{SC}(\{A,B\}) = 1-\frac{3}{8} = \frac{5}{8}$. The \rew{three} remaining edges are symmetric to $\{A,B\}$ and also have $\textsc{SC}=\frac{5}{8}$.}

In this work, we will be focusing on the problem of spanning centrality computation for all edges, i.e., {\em All Edges Spanning Centrality}~({\sc AESC}). The current state-of-the-art~\cite{teixeira2013spanning} for {\sc AESC} is based on Kirchhoff's matrix-tree theory and exhibits a time complexity of $\mathcal{O}(mn^{3/2})$, which is infeasible for analyzing large graphs. To overcome the computational burden, many approximation methods, based on random walks, have been proposed~ \cite{aesckdd,peng2021local,hayashi2016efficient,mavroforakis2015spanning}. In the approximation setting, given the graph $G = (V, E)$, an $\epsilon$-approximation algorithm outputs $\hat{\textsc{SC}}(e)$ for each $e \in E$ such that
 $$|\hat{\textsc{SC}}(e) - {\textsc{SC}}(e)| \leq \epsilon,~~\forall e \in E$$ where $\epsilon$ is an upper bound on the errors of reported \textsc{SC} values for all the edges. \rew{An output satisfying the above condition is called an
$\epsilon$-approximation. For randomized algorithms, the probability
of satisfying this condition is stated separately. }

\subsection{Random Walks}
A random walk in a graph $G = (V,E)$ can be defined as a sequence of vertices from $V$ in which each consecutive vertex pair is connected via an edge in $E$. In our setting, the walk $W = (v_0, v_1, \cdots, v_t)$ starts from a given, non-random vertex $v_0 \in V$, and each subsequent vertex $v_i$, $1 \leq i \leq t$ is chosen uniformly at random from the neighbors of $v_{i-1}$, that is $v_i \in \Gamma(v_{i-1})$ for all $1 \leq i \leq t$. \rew{A conventional neighbor selection is:}
    \begin{equation}
        \label{eq:select-naivest}
        \rew{v_i = \Gamma(v_{i-1})\left[{\cal R}~\bmod~d(v_{i-1})\right]}
    \end{equation}
where ${\cal R}$ is a large random integer. That is, the \rew{selection} first computes a random number between 0 and $d(v_{i-1}) - 1$ and returns the vertex with this index from  $\Gamma(v_{i-1})$.

The traditional random walk generation is given in Algorithm~\ref{algo:randomwalk}: first, the algorithm sets the starting vertex of the walk $W$ and the current vertex $cur$ to $v$. Subsequently, \rew{it performs $L$ transitions, or hops, so the conceptual path contains $L+1$ vertices including the starting vertex. At each step $1\leq l\leq L$}, it randomly selects one of $cur$'s neighbors and moves to that vertex using a random choice function at lines 4--5. The algorithm stores each visited vertex during the walk inside $W$ to track the path. In AESC, this process is repeated for $n_{req}$ times to generate multiple random walks starting from the same vertex~(line~\ref{ref:loop} of Algorithm~\ref{algo:tgtp}).

\renewcommand{\baselinestretch}{1.2}
\begin{algorithm}[!ht]
\small
\caption{\sc{RandomWalk}}
\label{algo:randomwalk}
\algorithmicrequire{
$G = (V,E)$: the graph
\\\hspace*{7.1ex}$v \in V$: starting vertex
\\\hspace*{7.0ex}\rew{$L$: number of transitions/hops in the random walk}
}
\\\algorithmicensure{${W}$: random walk
}
\begin{algorithmic}[1]
    \State \rew{$W\leftarrow\mbox{ empty list of length }L+1$}
    \State {$W[0] \leftarrow cur \leftarrow  v$}
    \For{\rew{$l=1$ {\bf to} $L$}}
        \State \rew{$k \leftarrow \mathcal{R}_{W,l} \bmod d(cur)$} \Comment{fresh random draw for walk $W$} \label{line:rng}
        \State { $ cur \leftarrow \Gamma_{cur}[k]$ } \Comment{choose the $k$th vertex within $\Gamma_{cur}$}
        \State { $W[l] \leftarrow cur$} \label{line:pathgrow}
        \EndFor
    \State {{\bf return} $W$}
\end{algorithmic}
\end{algorithm}
\renewcommand{\baselinestretch}{1}

 \rew{We define streaming selections from the current per-walk state $\mathcal{X}_W$ and vertex $cur$. The function $f(\mathcal{X}_W,cur)$ returns a neighbor index; the state is initialized on a uniform grid and updated at every streaming step (Section~\ref{sec:tasking}).}
\rew{In Algorithm~\ref{algo:randomwalk}, $\mathcal{R}_{W,l}$ denotes a fresh pseudo-random integer generated for walk $W$ at step $l$. The proposed bouquet method instead initializes per-walk states on a uniform grid and advances them together, as described below.}

\subsection{State-of-the-art in {\sc AESC} approximation: {\tt TGT+}}

Truncated Graph Traversal~(TGT), proposed by Zhang~et~al.~\cite{aesckdd}, overcomes the limitations of other existing methods, e.g.,~\cite{hayashi2016efficient,peng2021local,mavroforakis2015spanning}, by strategically limiting the number of vertices/edges visited via judiciously truncated traversals. Zhang et al. improved this algorithm in terms of both empirical efficiency and asymptotic performance while retaining result quality by proposing {\tt TGT+}, combining  {\tt TGT} with random walks and employing additional heuristic optimizations. In this work, we have used {\tt TGT+} as a real-world baseline to evaluate the performance of our random-walk processing scheme. We refer the reader to~\cite{aesckdd}, but for completeness of this work, the {\tt TGT+}  algorithm is described in Algorithm~\ref{algo:tgtp}. \rew{For connected, non-bipartite graphs and any $\epsilon,\delta\in(0,1)$, {\tt TGT+} returns an
$\epsilon$-approximation for all edges with probability at least $1-\delta$.}

\begin{algorithm}[!ht]
\renewcommand{\baselinestretch}{1.2}
\small
\caption{\sc{TGT+}}
\label{algo:tgtp}
\algorithmicrequire{$G = (V,E)$: the graph
\\\hspace*{6.8ex}$\epsilon$: The absolute error threshold
\\\hspace*{6.8ex}$\delta$: \rew{The failure probability}
\\\hspace*{6.8ex}$\omega$: The number of eigenvectors used
\\\hspace*{6.8ex}$\gamma$: \rew{The number of candidate vertices}
}
\\\algorithmicensure{\rew{$\hat{s}(e_{i,j})$}: \rew{Estimated spanning centrality of $e_{i,j}$,} $\forall e \in E$
}
\begin{algorithmic}[1]
    \For{$ v_i \in V $}\label{ln:loop}

        \State{$\forall v_j\in \Gamma(v_{i})~\tau_{i,j}\leftarrow $\sc{CalculateTau}$(e_{i,j}, \epsilon/2)$}
        \State {$p_0(u,v_i) \leftarrow 0~~\forall u \in {V}\setminus v_i $}
        \State{$p_0(v_i,v_i) \leftarrow 1$}
        \State {$\hat{g}_{\tau}(v_i,v_j) \leftarrow 1/d(v_i)~~~\forall v_j \in \Gamma(v_i)$}
        \State{\rew{$l\leftarrow 0$}}
        \State{$n_{walks} \leftarrow \sum_{v_j \in \Gamma(v_i)}$ {\sc NWalks}$(e_{i,j},\tau_{i,j})$}
        \State{\rew{$p_l(u,v_i) \leftarrow 0~~\forall u \in {V},~l \geq 1$}}
        \While{$\underset{u \in V \mbox { \& } p_l(u, v_i) \neq 0}{\sum} d(u)  < n_{walks}$
        }\label{algo:while}
            \For{\rew{$ v_j \in V \mbox{ s.t. } p_{l}(v_j,v_i)>0$}}
                \For{$v_x \in  \Gamma(v_j)$}
                    \State{\rew{$p_{l+1}(v_x,v_i)\leftarrow p_{l+1}(v_x,v_i) + \frac{p_{l}(v_j,v_i)}{d(v_{x})}$}}
                \EndFor
            \EndFor
            \State{$l\leftarrow l+1$}
            \For{\rew{$v_j\in\Gamma(v_i)\mbox{ s.t. }l\leq\tau_{i,j}$}}
                \State{\rew{$\hat{g}_\tau(v_i,v_j)\leftarrow\hat{g}_\tau(v_i,v_j)+\frac{p_l(v_i,v_i)-p_l(v_j,v_i)}{d(v_i)}$}}
            \EndFor
            \State{$n_{walks} \leftarrow \sum_{v_j \in \Gamma(v_i)}$ {\sc NWalks}$(e_{i,j}, \tau_{i,j}-l)$}
        \EndWhile
        \State{$\tilde{\tau} \leftarrow l$}
        \For{$v_j \in \Gamma(v_i) \mbox{ s.t. } \tau_{i,j}-\tilde{\tau}>0$}
            \State{\rew{$L\leftarrow\tau_{i,j}-\tilde{\tau}$}}
            \State{\rew{$\chi\leftarrow\textsc{CalculateChi}(G,v_i,v_j,p_{\tilde{\tau}}(\cdot,v_i),L,\gamma)$}}
            \State{\rew{$n_{req}\leftarrow\textsc{NWalks}(e_{i,j},L,\chi)$}}
                \For{$k = 1 \mbox{ to } n_{req}$} \label{ref:loop}
                    \State{\rew{$W_i\leftarrow\textsc{RandomWalk}(v_i,L)$}}
                    \State{\rew{$W_j\leftarrow\textsc{RandomWalk}(v_j,L)$}}
                    \State{\rew{$X\leftarrow\sum_{r=1}^{L}\bigl(p_{\tilde{\tau}}(W_i[r],v_i)-p_{\tilde{\tau}}(W_j[r],v_i)\bigr)$}}
                    \State {$\hat{g}_\tau(v_i,v_j) \leftarrow \frac{X}{n_{req}  d(v_i)}+\hat{g}_\tau(v_i,v_j) $}\label{ln:critical}
                \EndFor
        \EndFor
    \EndFor
    \For{ $e_{i,j} \in E$}
        \State{$\hat{s}(e_{i,j}) \leftarrow \hat{g}_\tau(v_i,v_j) + \hat{g}_\tau(v_j,v_i)$}
    \EndFor
    \State{\Return{ $\hat{s}(e_{i,j}) \rew{~\forall~e_{i,j}} \in E $}}
\end{algorithmic}
\renewcommand{\baselinestretch}{1}
\end{algorithm}
In a nutshell, {\tt{TGT+}} goes over each vertex $v_i \in V$~(the {\bf{for}} loop at line 1) and processes in two main steps; \rew{after initialization, the {\bf while} loop propagates transition probabilities and accumulates their deterministic contributions to $\hat{g}_\tau$, stopping at hop $\tilde{\tau}$.} The reasoning is that the number of vertices near, i.e., within the $\tilde{\tau}$-neighborhood, of $v_i$ is typically small, allowing for efficient coverage through a (breadth-first) graph traversal starting from $v_i$. In contrast, vertices that are farther away from $v_i$ can be numerous, potentially in the order of millions in large graphs. In such cases, random walks are more suitable, as they can prioritize exploring important vertices rather than attempting to cover all distant vertices. That is, the algorithm assumes the processing of the $\tilde{\tau}$-neighborhood of $v_i$ to be cheaper compared to a random-walk-based simulation.

The second step~(\rew{the {\bf for} loop over neighbors with positive remaining walk length}) iterates $n_{req}$ times per such neighbor, where $n_{req}$ is the required number of two-way random walks to reach sufficient accuracy. At each iteration, this step generates two random walks $W_i$ and $W_j$, starting from $v_i$ and $v_j$, respectively. Then the vertices on these length-$(\tau_{i,j} - \tilde{\tau})$ random walks $W_i$ and $W_j$ are used to improve edge centrality estimation where $\tau_{i,j}$ is the truncated length for each edge $\{v_i, v_j\} \in E$ computed via eigenvalues and eigenvectors of the matrix $\mathbf{D}^{1/2}\mathbf{P}\mathbf{D}^{-1/2}$. Here, $\mathbf{D}$ is the diagonal, degree matrix and $\mathbf{P} = \mathbf{D}^{-1}\mathbf{A}$ is the random-walk transition probability matrix for a given $G$ with the adjacency matrix $\mathbf{A}$. More details and the exact definitions of the {\sc{CalculateTau}}, {\sc{NWalks}}, and {\sc{CalculateChi}} functions can be found in the original work~\cite{aesckdd}. \rew{Because the random-walk component is our focus, it is given in Algorithm~\ref{algo:randomwalk}.}
\rew{Here, \textsc{CalculateChi} bounds the residual walk sum for length $L$. \textsc{NWalks} uses this bound $\chi$ in Eq.~(13) of~\cite{aesckdd}, rounding the sample count upward. Its two-argument calls use the coarse bound $2L(\max_u p_l(u,v_i)-\min_u p_l(u,v_i))$ at the current frontier, and return zero for nonpositive remaining lengths. The residual sums count hops $1,\ldots,L$, including repeated visits and excluding the starting vertex.}

\subsection{Hash-based Random Number Generation}
\rew{Pseudo-random number generators produce sequences by updating an internal state. Hash-based sampling combines a simulation seed with a hash of the graph element being sampled. In the edge-sampling method of G\"{o}kt\"{u}rk~and~Kaya~\cite{gokturk2020boosting}, a seed $\mathcal{X}_r$ identifies the $r$th simulation, and each edge $\{u,v\}\in E$ receives the value}
    \begin{equation}
        \label{eq:hash_prob}
        P(\{u,v\})_r = \frac{\mathcal{X}_r \oplus h(u,v)}{h_{max}},
    \end{equation}
where $P(\{u,v\})_r$ is the pseudo-random number compared against the sampling probability of $\{u,v\}$ while generating the $r$th sample. In the equation, $h(u,v)$ is the hash value of the edge $\{u,v\}$, $\mathcal{X}_r$ is an \rew{independently chosen} random seed for the $r$th sample, and \rew{$h_{max}$ denotes the size of the hash domain, so hash values lie in $\{0,\ldots,h_{max}-1\}$; throughout this work, $h_{max}=2^{32}$}.

\rew{For random walks, we retain this seed-and-hash composition but use the current vertex hash and an evolving per-walk state. Section~\ref{sec:tasking} gives the neighbor-selection rule and degree-based state update.}

\subsection{Vectorized Processing and SIMD}
{\em Single Instruction Multiple Data}~(SIMD) is a parallelization technique that enables a single instruction to perform the same operation on multiple data items concurrently. SIMD units are available in almost all modern CPUs and GPUs to achieve computational efficiency for tasks allowing data-level parallelism. In these units, the data is often organized into vectors, and a single instruction is applied to all elements in the vector. Modern processors often feature SIMD instruction sets such as Intel's AVX2~(Advanced Vector Extensions 2) or ARM's NEON.
These instructions typically improve the throughput, though they may have slightly higher latency.

\rew{GPUs expose analogous units as Streaming Multiprocessors or Compute Units executing warps/wavefronts of 32--64 threads in lockstep; the bouquet arrangement extends naturally to such architectures.}

\section{Sampling-Aware Bouquets for Approximate {\sc{AESC}}}\label{sec:method}

In modern processors, when the computation and data layout incur highly irregular memory accesses, such as sparse matrix, graph, and sparse tensor operations, performance is drastically impacted. As in many random-walk-based Monte Carlo implementations on graphs, in AESC, each walk is processed individually, and each step in a walk iteratively samples only {\em a single} vertex to traverse. This yields irregular memory access patterns since the vertex neighborhoods are irregular by nature. However, for most Monte Carlo algorithms, it is possible to rearrange the order of operations so that many random walks are processed together. In this work, instead of handling the sampling and processing steps individually for a single random walk, we propose to {{\em fuse} them within the course of a randomized algorithm. As explained later, this yields a memory-efficient lockstep processing scheme, which can be used to {\em regularize} the computation. This regularization improves both the spatial and temporal locality of the operations performed and allows more efficient use of the computing resources.
When $n_{req}$ is large, which is the case for AESC especially if the desired approximation error is small, these walks share vertices which can be exploited via vectorization.

\subsection{SIMD Execution and its Locality Bottleneck}
\rew{Each walk maintains a streaming state $\mathcal{X}_W$, initialized on a uniform grid. At each streaming step, the state is XOR'ed with the hash $h(cur)$ of the current vertex to select a neighbor, and is then multiplied by the degree $d(cur)$ of the departure vertex, with 32-bit wraparound. Thus, the hash does not depend on the walk level. Revisiting a vertex can use a different state, although the recurrence does not guarantee that repeated selections or cycles are avoided. The initial grid excludes zero. If the state becomes zero, only the affected lane switches to conventional AVX2 sampling for its remaining hops, as specified in~\eqref{eq:stream-cutoff}. For fixed per-walk fallback draws, each labeled walk is determined by its initial state and starting vertex; Proposition~1 states the conditions under which reordering preserves its outcome.}

Although vectorization can significantly improve performance, with the traditional random-walk generation process, exploiting the CPU's vector capabilities is impossible since each step lengthens {\em a single walk} by one. In this work, although we have implemented our methods only using AVX2 intrinsics on the CPU, they can also be applied while processing many random walks on GPUs.

The basic random walk generation process given in Algorithm~\ref{algo:randomwalk}, by its design, is not SIMD-friendly and furthermore, incurs an unpredictable memory access pattern. In simple terms, the algorithm performs iterative random number generation and random neighbor selection. The state-of-the-art RNGs are fast and only require a few simple instructions like integer addition and exclusive-or. In addition, the (short) latency of these instructions can be hidden, which makes them negligible compared to the overall runtime. This leaves edge traversal, i.e., random memory access, as the major bottleneck for the performance.
Furthermore, the expensive modulus operation executed while selecting the new vertex has a long {\em latency} that cannot be hidden since its result is immediately required. Both problems can be remedied by first scheduling vectorized random walks into {\em bouquets} as a first step.

\subsection{SABA: Sampling-Aware Bouquet Arrangement}\label{sec:tasking}

A vectorized processing scheme alone does not solve the memory access problems of random-walk processing without a naive sampling. In addition to vectorization, {\em sampling-aware bouquet arrangement}~(SABA) arranges the random walks with respect to their potential memory accesses by manipulating and exploiting the underlying random number generation scheme. Let $cur \in V$ be the current vertex for a random walk $W$. The straightforward neighbor-vertex sampling, given in~\eqref{eq:select-naive}, generates a pseudo-random number and constrains the result within the cardinality of the current vertex's neighborhood using the {\em modulus} operator, which induces a significant latency.
    \begin{equation}
        \label{eq:select-naive}
        \rew{k_{W,l} = \mathcal{R}_{W,l}~\bmod~d(cur)}
    \end{equation}
\noindent
Figure~\ref{fig:pathexample_bad} shows the case where 8 random walks, $W_0$ to $W_7$, are generated one after another. Algorithm~\ref{algo:randomwalk}, {\sc RandomWalk}, is called 8 times to generate these walks. Vectorized execution can also be employed without any arrangement: Figure~\ref{fig:vector_bad} shows the generation of the same set of walks in batches of size four, advancing four walks in lockstep with $B = 4$. For the first batch, the starting vertices of $W_0$ to $W_3$ are set, and their second vertices are randomly chosen at once. Lastly, their final vertices are chosen. Then the execution continues with the next batch, i.e., the walks $W_4$ to $W_7$.

\begin{figure}
     \centering
     \begin{subfigure}[b]{0.50\textwidth}
         \centering
         \includegraphics[width=\textwidth]{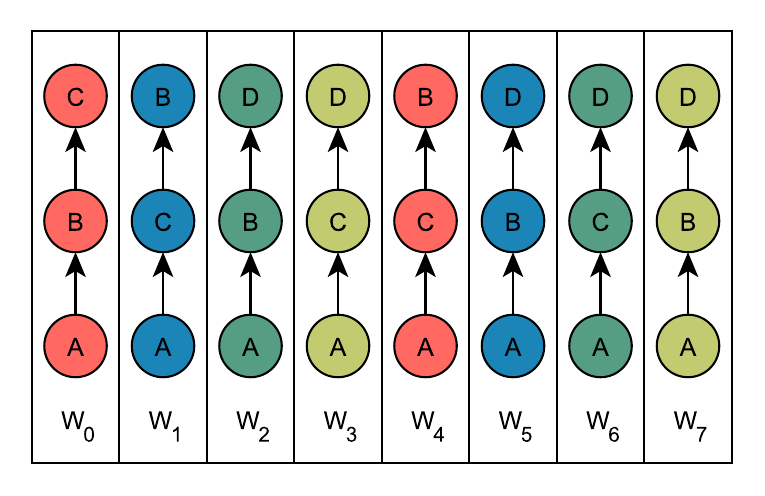}
         \caption{\small Sequential, non-vectorized processing of the walks in the order of their IDs.}
        \label{fig:pathexample_bad}
     \end{subfigure}
     \hfill
     \begin{subfigure}[b]{0.45\textwidth}
         \centering
         \includegraphics[width=\textwidth]{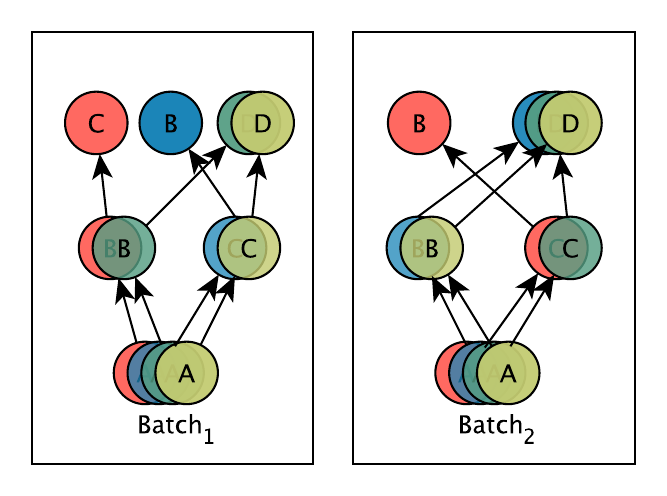}
         \caption{\small Vectorized processing of the walks with SIMD, four random walks at once.}
         \label{fig:vector_bad}
     \end{subfigure}
        \caption{A set of random walks generated from the sample graph in Fig.~\ref{fig:toy} with no judicious arrangement (left) and their vectorized execution (right).}
        \label{fig:badexample}
\end{figure}
Processing multiple random walks indeed allows us to perform vectorized operations for vertex selection and random number generation. However, when the corresponding memory accesses require different cache lines as in Fig.~\ref{fig:vector_bad}, the performance of the memory subsystem becomes the bottleneck. The proposed approach seeks to remedy this by sampling similar paths together. The computation is regularized by
\begin{enumerate}
\item \rew{a per-walk state $\mathcal{X}_W$, initialized on a uniform grid before execution and advanced by degree multiplication, and}
\item \rew{the hash $h(cur)$ of the current vertex.}
\end{enumerate}
{\em Sampling-aware bouquet arrangement}~(SABA) uses the hash-based RNG scheme below and \rew{arranges the walks within bouquets by ordering their initial grid states $\mathcal{X}_W$;}
\begin{equation}
    \label{eq:select-proposed}
    \rew{f(\mathcal{X}_W,cur)  = \left\lfloor \frac{{\mathcal X}_W \oplus h(cur) }  {h_{max}} \times d(cur) \right\rfloor}
\end{equation}
\rew{where $h_{max}=2^{32}$ and $0 \leq f(\mathcal{X}_W,cur)<d(cur)$. After selecting the neighbor, we stream the state forward using}
\begin{equation}
    \label{eq:state-stream}
    \rew{\mathcal{X}_W \leftarrow \bigl(\mathcal{X}_W \times d(cur)\bigr) \bmod h_{max}.}
\end{equation}
\rew{Here $cur$ denotes the departure vertex: its degree must be retained before advancing to the selected neighbor. The initial grid value is attached to the walk descriptor, whereas $\mathcal{X}_W$ evolves during execution. The state update uses the low 32 bits of the product; neighbor selection uses the high 32 bits of $(\mathcal{X}_W\oplus h(cur))d(cur)$.}
\rew{Before each hop, each lane checks its streaming state. A nonzero state uses the streaming selection and update above; a zero state uses a fresh pseudo-random draw as in~\eqref{eq:select-naive}:}
\begin{equation}
    \label{eq:stream-cutoff}
    \rew{k_{W,l}=\begin{cases}
        f(\mathcal{X}_W,cur), & \mathcal{X}_W\neq0,\\
        \mathcal{R}_{W,l}\bmod d(cur), & \mathcal{X}_W=0.
    \end{cases}}
\end{equation}
\rew{Initial grid values exclude zero, so every lane starts in streaming mode. If a streaming update makes $\mathcal{X}_W=0$, that lane switches to conventional AVX2 sampling from the next hop onward, using the same fresh pseudo-random draws and neighbor-selection rule as the Naive method. Its state remains zero, and it uses fallback for the remainder of the walk. Other lanes continue streaming while their own states are nonzero.}

\rew{\begin{proposition}[Execution Schedule Invariance]
\rew{Let $S=\{s_1,s_2,\ldots,s_N\}$ denote the fixed set of simulation identifiers. Each simulation $s\in S$ defines a deterministic outcome $X(s)$, with its initial grid value, starting vertex, and any fallback random draws fixed independently of the execution schedule. The Monte Carlo estimator is}
\[
\hat{\mu}=\frac{1}{N}\sum_{s\in S}X(s).
\]
Let $\pi:S\rightarrow S$ be an arbitrary permutation representing the execution schedule. Since $\pi$ is bijective, every simulation identifier appears exactly once in both the original and reordered schedules. Therefore,
\[
\sum_{s\in S}X(\pi(s)) = \sum_{s\in S}X(s),
\]
which immediately implies
\[
\hat{\mu}_{\pi}=\hat{\mu}.
\]
\end{proposition}
}

\rew{
\rew{A simulation identifier corresponds to a labeled walk with its initial grid value and any fallback random draws attached to that identity. Reordering preserves the finite set of walk outcomes when these inputs remain fixed. In particular, fallback draws must follow the walk identity for the pointwise equality in Proposition~1 to apply; assigning draws solely by their consumption order in a shared generator does not establish that equality. The proposition concerns execution order and does not establish independent sampling or the estimator error bound for uniform-grid initialization.}
}

\rew{
\begin{theorem}[Range-Reduction Bias]
\rew{For a fixed degree $d>0$, let $U$ be uniform over $\{0,\ldots,h_{max}-1\}$ and let $K=\lfloor Ud/h_{max}\rfloor$. For any edge index $k$, $0\leq k<d$, we have}
\[
\rew{\Pr[K=k] =}
\frac{\#\{u:\; k h_{max}/d \le u < (k{+}1) h_{max}/d\}}{h_{max}},
\]
where the count in the numerator is either $\lfloor h_{max}/d \rfloor$ or $\lceil h_{max}/d \rceil$. Therefore,
\[
\rew{\max_{0 \le k < d}\left|\Pr[K=k] - \frac{1}{d}\right|}
~\le~ \frac{1}{h_{max}}.
\]
\end{theorem}
}

\rew{
\rew{This theorem isolates range-reduction bias under uniform input. It does not establish that the uniform-grid initialization or the subsequent streaming states provide such input. Multiplication modulo $2^{32}$ need not preserve a uniform state distribution. A lane switches to Naive sampling when its streaming state becomes zero, as in~\eqref{eq:stream-cutoff}; this safeguard does not itself prove uniformity during streaming. Agreement with the reference implementation is assessed empirically in Section~\ref{sec:aesc-exp}. Under the same uniform-input assumption, modulus-based range reduction also has a per-neighbor deviation of at most $1/h_{max}$.}
}
\rew{For fixed initial states and fixed per-walk fallback draws, Proposition~1 gives the same estimator under any permutation of the execution order. This statement applies to the complete sampling scheme and does not imply equivalence to a different random-walk generator.}
\rew{The comparison with the original {\tt TGT+} implementation is empirical: Section~\ref{sec:aesc-exp} compares their mean absolute errors against the same reference centralities.}

\rew{The main benefit of SABA is that constructing bouquets from consecutive grid values places walks with nearby initial states together. Nearby states often share a high-order bit prefix. When lanes have the same current vertex, XOR with the common value $h(cur)$ preserves that shared prefix. The contiguous range reduction in~\eqref{eq:select-proposed} then tends to map those lanes to the same or nearby adjacency indices. This locality effect is strongest at the first steps. Degree multiplication can change shared prefixes, and lanes reaching different vertices use different hashes. A lane whose state becomes zero switches to conventional AVX2 sampling, which no longer uses its ordered streaming state. Other lanes retain their streaming states until they also become zero or complete their walks. Later-step locality is therefore a graph-dependent heuristic, evaluated through the cache and branching measurements in Section~\ref{sec:cache}. The schedule-preservation conditions, including fallback draws, are stated in Proposition~1.}

\renewcommand{\baselinestretch}{1.2}
\begin{algorithm}[!ht]
\small
\caption{\sc{RandomBouquet}}
\label{algo:randomwalk-simd}
\algorithmicrequire
$v \in V$: starting vertex
\\\hspace*{6.7ex}$k$: number of walk descriptors to generate
\\\hspace*{6.7ex}$B$: SIMD vector size/\#random walks per bouquet
\\\hspace*{6.7ex}\rew{$L$: number of transitions/hops in each walk}\\
\algorithmicensure{$\{W_{0},\dots,W_{k-1} \}$: $k$ random walks arranged into bouquets}

\begin{algorithmic}[1]
    \For{$j = 0$ {\bf to} $k-1$}
  \State \rew{$\mathcal{X}_{W_j} \leftarrow \textsc{UniformGrid}(j,k)$; $W_j \leftarrow$ empty list of length $L+1$}
    \EndFor
    \State \rew{form bouquets from consecutive descriptors in the ascending grid order}
    \For{each consecutive group $W_b,\dots,W_{b+B-1}$ of $B$ descriptors}
        \State{$W_r[0] \leftarrow cur_r \leftarrow v$, for $b \leq r < b+B$} \label{lnlnln}
        \For{\rew{$l=1$ {\bf to} $L$}} \label{ln:bouquetstep}
            \State \rew{$\mathcal{A}\leftarrow\{r:b\leq r<b+B,\ \mathcal{X}_{W_r}\neq0\}$}
            \State \rew{$k_r\leftarrow\lfloor((\mathcal{X}_{W_r}\oplus h(cur_r))/h_{max})d(cur_r)\rfloor$, for $r\in\mathcal{A}$} \label{line:rngvec}
            \State \rew{$\mathcal{X}_{W_r}\leftarrow(\mathcal{X}_{W_r}d(cur_r))\bmod h_{max}$, for $r\in\mathcal{A}$}
            \State \rew{draw fresh $\mathcal{R}_{W_r,l}$ and set $k_r\leftarrow\mathcal{R}_{W_r,l}\bmod d(cur_r)$, for $b\leq r<b+B$ with $r\notin\mathcal{A}$}
            \State { $cur_{r} \leftarrow \Gamma_{cur_{r}}[k_{r}]$, for $b \leq r < b+B$  }
            \State { $W_{r}[l] \leftarrow cur_{r}$, for $b \leq r < b+B$  }
        \EndFor
    \EndFor
    \State \bf{return} {$\{ W_{r}:~ 0 \leq r < k$\} }
\end{algorithmic}
\end{algorithm}
\renewcommand{\baselinestretch}{1}
\rew{For $k$ walks, the initial states are positive multiples of a fixed integer spacing:}
\begin{equation}
    \label{eq:uniform-grid}
    \rew{\textsc{UniformGrid}(j,k)=(j+1)\left\lfloor\frac{2^{32}-1}{k}\right\rfloor,\qquad 0\leq j<k.}
\end{equation}
\rew{For $1\leq k\leq2^{32}-1$, these states are nonzero, distinct, and in ascending order. Algorithm~\ref{algo:randomwalk-simd} therefore constructs ordered bouquets directly. Each walk performs exactly $L$ hops. At the start of each hop, $\mathcal{A}$ identifies the lanes with nonzero states. These lanes use streaming; the remaining lanes use fresh Naive-method draws. The set $\mathcal{A}$ is fixed for that hop before the state updates, so a lane whose update produces zero starts fallback at the following hop. SIMD operations apply to the corresponding lane subsets.}
\rew{Algorithm~\ref{algo:randomwalk-simd} presents a conceptual path-materializing formulation. In the actual implementation, each visited vertex is consumed immediately by the accumulation operation, and complete walk paths are not stored. The number of walks in this work is always a multiple of $B$, \rew{so every bouquet is full; no lanes are disabled to handle a partial final bouquet.}}

\begin{figure}
     \centering
     \begin{subfigure}[b]{0.48\textwidth}
         \centering
         \includegraphics[width=\textwidth]{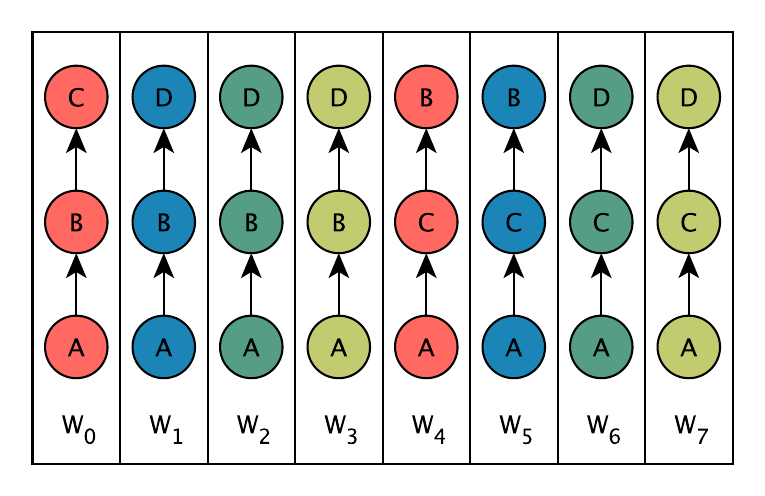}
         \caption{\small Sequential, non-vectorized processing of the walks \rew{in the order of their initial $\mathcal{X}$ values.}}
        \label{fig:pathexample}
     \end{subfigure}
     \hfill
     \begin{subfigure}[b]{0.42\textwidth}
         \centering
         \includegraphics[width=\textwidth]{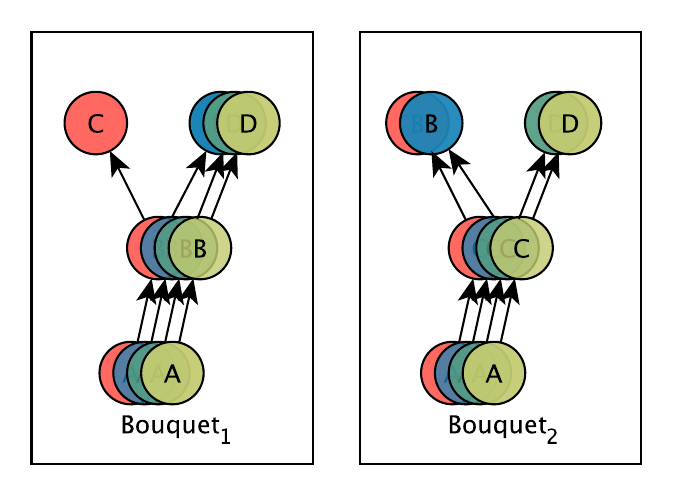}
         \caption{Vectorized processing of the walks with a SIMD unit. Each SIMD unit processes four random walks.}
         \label{fig:bouquet}
     \end{subfigure}
        \caption{\small A set of random walks generated from the sample graph in Figure~\ref{fig:toy} with SABA (left) and their vectorized execution (right). \rew{The figure illustrates walks arranged by their initial uniform-grid values.}}
        \label{fig:goodexample}
\end{figure}

Without SABA, as in Fig.~\ref{fig:badexample}, the first batch starts with $A$ and follows $B$ and $C$ from $A$'s neighborhood. Depending on the distance of $B$ and $C$ in memory, at most two cache lines may be read. In the second step, $C$ and $D$ (from $\Gamma(B)$) and $B$ and $D$ (from $\Gamma(C)$) are chosen. SABA aims to \rew{reduce} the number of distinct vertices at each step.
Indeed, when SABA reorganizes the walks as in Figure~\ref{fig:goodexample}, \rew{only one vertex is accessed in the first step of each batch: $B$ for the first batch and $C$ for the second}. \rew{Similarly, in the second step of both batches, two vertices are accessed instead of three.}

\section{Implementation Details}\label{sec:impdetail}

\rew{Parallelizing the original sequential {\tt TGT+} implementation~\cite{aesckdd} requires synchronizing its shared runtime accumulators and edge-keyed containers, which do not admit concurrent writers.}
\rew{The outer source loop of Algorithm~\ref{algo:tgtp} admits parallel execution. Both the deterministic and sampled contributions for source $v_i$ update its directed quantities $\hat{g}_\tau(v_i,v_j)$, whose entries are distinct from those of other sources. If contributions from both endpoints are accumulated directly into a shared undirected edge-result slot, concurrent updates to that slot require synchronization.}

\rew{In our implementation, each source vertex is assigned to one worker, giving that worker exclusive ownership of its source-local computation. Shared edge-result updates use atomic accumulation, and workers do not keep private copies of the complete $\hat{g}$ array. The extra storage per thread consists of scratch buffers of size $\Theta(n)$; hence, the overall footprint is $\Theta(m + T\cdot n)$ for $T$ threads.}
\rew{Each walk descriptor occupies $8$ bytes, giving $\Theta(n_{req})$ descriptors per source, and a bouquet holds $\Theta(B)$ words of transient state during processing; walk paths are never materialized, so this footprint is independent of $L$.}
The original {\tt{TGT+}} implementation relies on sparse data structures, specifically the C++ {\tt std::map}, to store edge-specific information. Swapping compilers and tweaking compilation flags can significantly increase its performance. Instead of relying on compiler optimizations, in our implementation, we have refrained from using these data structures. Wherever intermediate data structures are used, e.g., the {\tt map} with edges as keys in the original {\tt{TGT+}} implementation, we have skipped these steps and update the single shared result array in place.

\rew{In SABA and other vectorized baselines, we have used explicit AVX2 intrinsics to exploit instruction-level parallelism for hashing, XOR, and multiplication operations for generating random numbers. As the hash function, we used a vectorized implementation of MurmurHash3's 32-bit finalizer, \texttt{fmix32}~\cite{murmur3}. \rew{Neighbor selection in~\eqref{eq:select-proposed} uses the high 32 bits of an unsigned 32-by-32-bit product. The state update in~\eqref{eq:state-stream} retains the low 32 bits of the state-degree product. Both streaming operations avoid integer division and modulus instructions.}}
\rew{In the AESC experiments, Baseline includes the data-structure and memory optimizations of our {\tt TGT+} reimplementation. SABA adds uniform-grid streaming, per-lane fallback, and ordered bouquets as a combined sampling and scheduling design. These components are designed to work together, and the Baseline/SABA ratio measures their joint benefit beyond the shared optimizations.}

\rew{The scalar and conventional SIMD walk kernels use the C++ {\tt std::mt19937} implementation of Mersenne Twister~\cite{matsumoto1998mersenne} for their random draws. SABA uses the streaming state and vertex hash defined in Section~\ref{sec:tasking}, with conventional AVX2 sampling for a lane once its streaming state becomes zero, as in~\eqref{eq:stream-cutoff}. Fallback uses the same fresh pseudo-random draws and modulo-based neighbor selection as the Naive random-walk method in Algorithm~\ref{algo:randomwalk}; it does not introduce a separate generator.}

\section{Experimental Results}\label{sec:exps}

The experiments are conducted on \rew{a dual-socket server with two {\tt
Intel(R) Xeon(R) E5-2620v4} CPUs~(16 cores in total)}, \rew{with a base frequency of
2.10GHz and approximately $189$\,GiB of usable memory}. \rew{Each processor provides a $20$\,MB shared last-level (L3) cache; each core provides $256$\,KB of private L2 cache and $32$\,KB of private L1 data cache. The reported miss counts correspond to misses at the L3 level, collected with {\tt perf stat} using the {\tt LLC-misses} hardware event while the walk-computation phase runs, with parsing and pre-processing excluded as for all reported times.} \rew{Each processor provides four DDR4 memory channels. Intel MLC v3.12 measured an idle memory latency of $82$\,ns and a peak read bandwidth of $92.4$\,GB/s over both sockets.} The operating system on the server is {\tt
Ubuntu Linux
} with
5.4.0-167
kernel. \rew{The CPU algorithms are implemented using {\tt C++20} and compiled with {\tt Intel C++ Compiler 2023.2.1}. All compared implementations, including the original single-threaded {\tt TGT+}, use the same compiler and optimization flags ({\tt -Ofast}, {\tt -march=native}). Multithreading uses OpenMP, and AVX2 operations use explicit intrinsics.}
\rew{OpenMP threads were bound to core places using
\texttt{OMP\_PLACES=cores} and \texttt{OMP\_PROC\_BIND=close}; OpenMP thread counts were capped at the values reported for each experiment.}

\begin{table}[!ht]
\renewcommand{\arraystretch}{1.2}
\caption{Properties of networks used in the experiments}\label{tab:NetProps}
\centering

\begin{tabular}{l||r|r|r}
& No. of           & No. of    & \\
Dataset & vertices          & edges           & Diameter               \\
\hline
{\tt California} \rew{(roadNet-CA)} & 1,965,206 & 2,766,607 & 865\\
{\tt Delaunay-N21} \rew{(delaunay\_n21)} & 2,097,152 & 6,291,408 & 623\\
{\tt Epinions} \rew{(soc-Epinions1)} & 75,879 & \rew{405,739} & 14 \\
{\tt Facebook} \rew{(facebook\_combined)} & 4,039 & \rew{88,233} & 8\\
{\tt Gnutella} \rew{(p2p-Gnutella31)} & 62,586 & \rew{147,891} & 11 \\
{\tt HepPh} \rew{(cit-HepPh)} & 34,546 & \rew{420,920} & 12\\
{\tt HepTh} \rew{(cit-HepTh)} & \rew{27,770} & \rew{352,323} & 13 \\
{\tt Orkut} \rew{(com-Orkut)} & 3,072,441 & 117,185,083 & 9 \\
{\tt Slashdot} \rew{(soc-Slashdot0902)} & 82,168 & \rew{582,532} & 11 \\
{\tt Twitch} \rew{(twitch\_gamers)} & \rew{168,114} & \rew{6,797,557} & \rew{8} \\
\end{tabular}

\renewcommand{\arraystretch}{1}
\end{table}

The experiments are performed on ten graphs given in Table~\ref{tab:NetProps}.
\rew{{\tt California} is the road network of the state of California, where intersections and road segments between them form the vertices and edges,}
\rew{{\tt Delaunay-N21} is a Delaunay triangulation mesh generated over randomly sampled points in the unit square,}
{\tt Epinions} is a trust network for the consumer review site Epinions,
{\tt Facebook} is the friendship network collected from survey participants for various apps, {\tt Gnutella} is a snapshot of the Gnutella peer-to-peer file sharing network,
\rew{{\tt HepPh} and {\tt HepTh} are the citation networks of the arXiv High Energy Physics Phenomenology and Theory categories, respectively,}
{\tt Orkut} is the friendship network of the social network Orkut,
{\tt Slashdot} is the friend/foe network of the users of Slashdot, a tech-related website, and \rew{{\tt Twitch} is the follower network of Twitch users provided as the Twitch Gamers dataset}.
\rew{All graphs except {\tt Delaunay-N21}, which is taken from the 10th DIMACS Implementation Challenge via the Network Data Repository,} are retrieved from Stanford Large Network Dataset Collection~\cite{snapnets}.
\rew{The directed networks were converted to undirected graphs by
removing the edge orientation. Preprocessing removes self-loops, duplicate
edges, and isolated vertices, so every retained vertex has positive degree.
Table~\ref{tab:NetProps} reports the processed graph sizes.}
\rew{For disconnected inputs, each remaining connected component is processed independently, since neither the truncated traversals nor the random walks can leave their own component.} On these datasets, we design two benchmark settings for a comprehensive experimental evaluation.

For each network dataset, we perform two sets of experiments;
\begin{enumerate}
  \setlength{\itemsep}{0.2pt}
  \setlength{\parskip}{0pt}
  \setlength{\parsep}{0pt}
    \item {\em Synthetic benchmark}: These benchmarks are done with the parameter sets;
    \begin{itemize}
     \item {Number of transitions/hops per walk: $L\in\{5,10,15\}$.}
     \item {Number of random walks per vertex: \{2048, 16,384\}.}
    \end{itemize}
    \item {\em AESC benchmark}: These benchmarks are done with a performance comparison to {\tt TGT+} by using the error parameters $\epsilon = 0.005$, $0.01$, and $0.05$; $\epsilon=0.05$ is the default error parameter from~\cite{aesckdd}. \rew{Following~\cite{aesckdd}, all AESC experiments use $\delta=1/n$, $\omega=128$ eigenvectors, and $\gamma=10$ candidate vertices.}
\end{enumerate}

\noindent With these benchmarks, we measure performance gains obtained via SABA on both real-world AESC computations where many, usually short, random walks per vertex are processed, and on other use cases where a large number of walks are processed as in our synthetic benchmarks.
In all these, we measure the total time spent on the actual computation, excluding parsing the datasets and pre-processing. \rew{The one-time offline eigendecomposition-based preprocessing of {\tt TGT+} is excluded identically from every implementation.} \rew{Every reported time is the best of three repeated runs.} \rew{All SABA timings include uniform-grid initialization and bouquet construction.}
\rew{For the synthetic benchmarks, visited-vertex counts consume the walk outputs during traversal. The reported kernel timings include traversal and counter updates; graph parsing and preprocessing are excluded as stated above. Complete paths are not stored.}

\subsection{Performance on AESC Benchmark}\label{sec:aesc-exp}

\rew{We compared SABA with the original {\tt TGT+} implementation on Facebook, HepTh, HepPh, and Gnutella at $\epsilon\in\{0.005,0.01,0.05\}$, with one run per implementation and setting. Both implementations had identical mean absolute errors against the same exact centralities to four decimal places in all twelve configurations.}

\begin{table*}[htbp]
\caption{Execution times (in secs, best of repeated runs) on the AESC benchmark for the original single-threaded {\tt TGT+}, our reimplementation of it~({\tt Baseline}), \rew{and our variant using uniform-grid streaming, per-lane fallback, and ordered bouquets~({\tt SABA})}. \rew{The Orkut and Delaunay-N21 experiments did not run to completion within three days and are therefore omitted from this table.}\label{tab:varw}}
\centering
\renewcommand{\tabcolsep}{2pt}
\resizebox{\textwidth}{!}{
\begin{tabular}{l|r||r|r|r|r|r|r|r}
\hline
 & Method & {\texttt{TGT+}} & {\texttt{Baseline}} & \multicolumn{5}{c}{\texttt{SABA}} \\ \hline
~Dataset & $\epsilon$/\#Threads & 1 & 1 & 1 & 2 & 4 & 8 & 16 \\ \hline
\multirow[c]{3}{*}{\texttt{California}} & 0.005 & 10025.70 & 655.65 & 586.94 & 385.80 & 643.50 & 419.38 & 210.28 \\
 & 0.010 & 10024.63 & 844.24 & 584.25 & 386.71 & 643.94 & 419.41 & 210.60 \\
 & 0.050 & 10081.56 & 948.52 & 584.00 & 388.16 & 643.40 & 419.38 & 210.31 \\ \cline{1-9}
\multirow[c]{3}{*}{\texttt{Epinions}} & 0.005 & 12.77 & 2.26 & 1.72 & 0.92 & 0.53 & 0.33 & 0.19 \\
 & 0.010 & 12.89 & 2.28 & 1.72 & 0.92 & 0.53 & 0.33 & 0.19 \\
 & 0.050 & 13.01 & 2.28 & 1.72 & 0.93 & 0.53 & 0.33 & 0.19 \\ \cline{1-9}
\multirow[c]{3}{*}{\texttt{Gnutella}} & 0.005 & 7.54 & 0.92 & 0.73 & 0.39 & 0.22 & 0.13 & 0.07 \\
 & 0.010 & 7.42 & 0.89 & 0.73 & 0.39 & 0.22 & 0.13 & 0.07 \\
 & 0.050 & 7.49 & 0.85 & 0.73 & 0.40 & 0.22 & 0.13 & 0.07 \\ \cline{1-9}
\multirow[c]{3}{*}{\texttt{Facebook}} & 0.005 & 2.82 & 1.70 & 1.14 & 0.56 & 0.31 & 0.19 & 0.10 \\
 & 0.010 & 2.19 & 0.90 & 0.60 & 0.30 & 0.17 & 0.10 & 0.06 \\
 & 0.050 & 1.66 & 0.28 & 0.20 & 0.10 & 0.06 & 0.04 & 0.02 \\ \cline{1-9}
\multirow[c]{3}{*}{\texttt{Twitch}} & 0.005 & 1497.28 & 1481.03 & 840.72 & 419.68 & 226.73 & 129.80 & 68.06 \\
 & 0.010 & 709.73 & 664.05 & 354.62 & 179.35 & 97.75 & 56.53 & 29.70 \\
 & 0.050 & 225.17 & 69.74 & 47.48 & 24.61 & 13.82 & 8.19 & 4.50 \\ \cline{1-9}
\multirow[c]{3}{*}{\texttt{HepPh}} & 0.005 & 7.43 & 1.04 & 0.78 & 0.41 & 0.24 & 0.14 & 0.07 \\
 & 0.010 & 7.44 & 1.06 & 0.77 & 0.41 & 0.23 & 0.14 & 0.07 \\
 & 0.050 & 7.46 & 1.11 & 0.77 & 0.41 & 0.23 & 0.14 & 0.07 \\ \cline{1-9}
\multirow[c]{3}{*}{\texttt{HepTh}} & 0.005 & 5.84 & 0.84 & 0.62 & 0.33 & 0.19 & 0.11 & 0.06 \\
 & 0.010 & 5.83 & 0.85 & 0.63 & 0.33 & 0.19 & 0.11 & 0.06 \\
 & 0.050 & 5.89 & 0.85 & 0.62 & 0.33 & 0.19 & 0.11 & 0.06 \\ \cline{1-9}
\multirow[c]{3}{*}{\texttt{Slashdot}} & 0.005 & 17354.64 & 17026.24 & 7307.05 & 3674.82 & 2001.68 & 1164.10 & 578.97 \\
 & 0.010 & 14019.36 & 13722.61 & 5806.58 & 2910.11 & 1590.66 & 920.56 & 462.56 \\
 & 0.050 & 7642.47 & 7956.13 & 2836.37 & 1417.81 & 775.67 & 448.86 & 223.64 \\ \cline{1-9}

 \multicolumn{2}{l|}{{Speedup vs {\tt TGT+}}} & $1.00\times$ & $4.34\times$ & $6.51\times$ & $12.13\times$ & $18.90\times$ & $31.08\times$ & $58.23\times$ \\
\multicolumn{2}{l|}{{Speedup vs 1-thread {\tt Baseline}}} & & \rew{$1.00\times$} & \rew{$1.50\times$} & \rew{$2.80\times$} & \rew{$4.36\times$} & \rew{$7.17\times$} & \rew{$13.42\times$} \\
\multicolumn{2}{l|}{{Speedup w.r.t. single thread}} & & & $1.00\times$ & $1.86\times$ & $2.90\times$ & $4.77\times$ & $8.94\times$ \\ \hline
\end{tabular}}
\end{table*}

\rew{Table~\ref{tab:varw} compares complete implementations: the original single-threaded {\tt TGT+}, our reference Baseline, and SABA using uniform-grid streaming with per-lane fallback and ordered bouquets. Fallback was rarely required in the AESC workload.}
\rew{Across the eight datasets and three error settings in Table~\ref{tab:varw}, the single-threaded Baseline and SABA pipelines are $4.34\times$ and $6.51\times$ faster than {\tt TGT+} on average (geometric mean). SABA is $1.50\times$ faster than Baseline.} We have elected not to include multithreaded results for the original {\tt TGT+} because its implementation is single-threaded.
\rew{A direct parallelization of the reference implementation does not produce a competitive multithreaded variant, since its shared runtime accumulators and edge-keyed maps would require synchronization across threads.} \rew{Nevertheless,} we report the speedup of our multi-threaded implementation with respect to {\tt TGT+} for completeness.

\FloatBarrier

\begin{figure}[!hbp]
    \centering
    \includegraphics[width=0.75\linewidth]{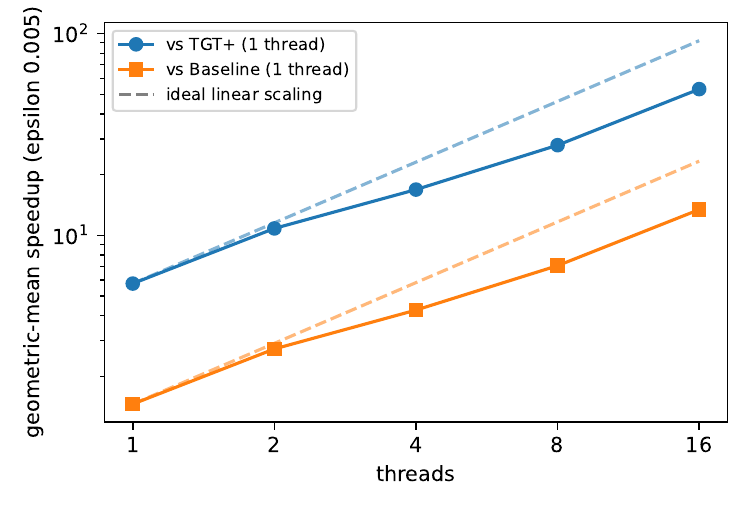}
    \caption{\rew{AESC speedup at $\epsilon=0.005$: geometric means over the eight completed datasets, relative to single-threaded {\tt TGT+} and single-threaded Baseline. Dashed lines show ideal linear scaling anchored at each single-thread result.}}
    \label{fig:aesc}
\end{figure}

\rew{With 16 threads, SABA is $58.23\times$ faster than single-threaded {\tt TGT+}, $13.42\times$ faster than single-threaded Baseline, and $8.94\times$ faster than its own single-thread execution. These aggregate comparisons combine the effects of the rewritten pipeline and parallel execution. The single-thread Baseline/SABA comparison is $1.50\times$.}
\rew{Figure~\ref{fig:aesc} shows strong scaling at $\epsilon=0.005$, using the eight completed datasets in Table~\ref{tab:varw}. This figure uses only that error setting; the table summary includes all three error settings.}
\rew{The California results show non-monotone scaling: at $\epsilon=0.005$, SABA takes $385.80$ seconds with two threads, $643.50$ seconds with four threads, and $210.28$ seconds with sixteen threads. Consequently, the aggregate speedups should not be read as linear scaling on every dataset. Orkut and Delaunay-N21 have no completed AESC entries in this table and are included only in the synthetic and cache experiments.}

\FloatBarrier

\begin{table}
\centering
\caption{\rew{Cache misses for 16,384 walks per vertex with $L=15$ hops. Naive reports absolute counts; AVX2 and SABA report percentages of Naive. Bottom rows give geometric-mean percentages relative to each baseline.}}
\label{tab:cachemisses}

\begin{tabular*}{\columnwidth}{@{\extracolsep{\fill}} l||rrr}
Dataset  & \multicolumn{1}{c}{Naive}         & \multicolumn{1}{c}{{AVX2}}     & \multicolumn{1}{c}{{SABA}}  \\
\hline
\texttt{California} & 446,529,490 & 55.33\% & 1.82\% \\
\texttt{Delaunay-N21} & 514,415,696 & 55.67\% & 3.12\% \\
{\tt Epinions} & 6,961,063,497    & 132.59\%   & 0.01\% \\
{\tt Facebook} & 845,487,783     & 134.95\%  & 0.01\% \\
{\tt Gnutella} & 4,244,404,537    & 127.55\%   & 0.01\% \\
{\tt HepPh}   & 2,891,186,237    & 127.86\%   & 0.01\% \\
{\tt HepTh}    & 2,286,294,504    & 133.06\%  & 0.01\% \\
{\tt Orkut}   & 1,077,204,257,701 & 142.17\%  & 69.03\% \\
{\tt Slashdot} & 7,210,297,453    & 119.48\%   & 0.02\% \\
{\tt Twitch}   & 41,381,402,038   & 114.69\%   & 57.37\% \\
\hline
\multicolumn{1}{l||}{Geo.\ mean vs {\tt Naive}} && $108.83\%$ & $0.16\%$\\
\multicolumn{1}{l||}{Geo.\ mean vs {\tt AVX2}} && & $0.15\%$\\
\end{tabular*}
\renewcommand{\arraystretch}{1}
\end{table}

\subsection{Effects of Bouquets on Cache}\label{sec:cache}

We have described {\em bouquets} as a technique to obtain pseudo-randomness while focusing on the performance; from another perspective, bouquets perform {\em best-effort}, {\em zero-shot spatial ordering} of walks.
Table~\ref{tab:cachemisses} shows that SABA with bouquets drastically reduces cache misses. As the third column of the table shows~(AVX2), naive vectorization significantly increases the memory pressure while processing multiple random walks; \rew{averaged geometrically over the graphs, direct vectorization yields $108.83\%$ of the naive method's misses, i.e., an $8.83\%$ increase.} For AVX2, 8 random numbers are generated in a non-vectorized manner and the modulo-based randomized selection, $$rnd(.)~\bmod~d(v),$$ is performed using AVX2 instructions.
As the last column shows, {SABA} clusters \rew{the same or nearby vertices} in the same vector and makes the consecutive memory accesses available in the cache. This approach reduces cache misses \rew{to $0.16\%$ of the naive method ($99.84\%$ reduction, geo. mean) and to $0.15\%$ of the AVX2 method's misses}.

\begin{table}[htbp]
\caption{Branching statistics for 8-wide random-walk vectors on the Facebook
dataset with 16,384 walks per source vertex. For each walk length, the table
reports the distribution of the number of distinct vertices in a vector,
pooled over all walk levels and sources.}
\centering
\renewcommand{\tabcolsep}{12pt}
\begin{tabular}{l|l||r|r|r|r|r}
\hline
~Length & Method & $1^{\mathrm{st}}$ & $10^{\mathrm{th}}$
       & $25^{\mathrm{th}}$ & $50^{\mathrm{th}}$ & Mean \\ \hline
\multirow[c]{2}{*}{5} & \texttt{AVX2} & 3 & 6 & 7 & 8 & 7.43 \\
 & \texttt{SABA} & 1 & 1 & 1 & 7 & 4.96 \\ \hline
\multirow[c]{2}{*}{10} & \texttt{AVX2} & 4 & 7 & 8 & 8 & 7.64 \\
 & \texttt{SABA} & 1 & 1 & 6 & 8 & 6.39 \\ \hline
\multirow[c]{2}{*}{15} & \texttt{AVX2} & 5 & 7 & 8 & 8 & 7.71 \\
 & \texttt{SABA} & 1 & 2 & 7 & 8 & 6.88 \\ \hline
\end{tabular}
\end{table}
\FloatBarrier

\rew{When lanes visit distinct vertices, their adjacency accesses spread across more memory locations. In the Facebook branching measurements, conventional AVX2 processing visits an average of $7.43$--$7.71$ distinct vertices per 8-wide vector across the tested walk lengths. SABA averages $4.96$, $6.39$, and $6.88$ distinct vertices at lengths $5$, $10$, and $15$, respectively. The increasing spread suggests that the locality benefit is strongest in the early steps. These measurements describe the scheduled lanes jointly; they do not establish the sampling distribution of an individual walk.}

\subsection{Performance on Synthetic Benchmarks}

\rew{The $6.51\times$ single-threaded speedup over {\tt TGT+} in Table~\ref{tab:varw} includes the data-structure and memory changes shared by Baseline and SABA. Their single-thread comparison gives a $1.50\times$ speedup.} We additionally assess random-walk processing through synthetic experiments with fixed walk counts and lengths.
\rew{In Algorithm~\ref{algo:tgtp}, the length of the walks generated for edge $(v_i,v_j)$ at a given level $\tilde{\tau}$ equals $\tau_{i,j}-\tilde{\tau}$ deterministically; the synthetic benchmarks vary precisely this length parameter, thereby isolating its effect on scheduling performance.} \rew{For each graph, we process the stated number of walks from each vertex for exactly $L$ transitions.} \rew{Each visited vertex is consumed by a counter update during the timed traversal, without storing complete paths.}

\begin{table*}
\caption{Execution times~(in secs) of the random walk generation/processing approaches mentioned in this work using 16 threads; the left group reports $2048$ walks per vertex and the right group $16{,}384$. \rew{All speedup rows are geometric means over the 30 dataset/length configurations.}}
\label{tab:rwtime}
\centering
\renewcommand{\tabcolsep}{2pt}
\resizebox{0.8\textwidth}{!}{
\begin{tabular}{lr||rrr||rrr}
Dataset & \rew{Hops} & \multicolumn{3}{c||}{\#Walks = 2048} & \multicolumn{3}{c}{\#Walks = 16,384} \\
& & Naive & AVX2 & SABA & Naive & AVX2 & SABA \\
\hline
\multirow[c]{3}{*}{\tt California}
& 5  & 61.10 & 32.26 & 4.76 & 456.52 & 227.53 & 38.03 \\
& 10 & 278.51 & 121.98 & 9.26 & 2197.37 & 945.33 & 74.01 \\
& 15 & 673.61 & 273.91 & 13.72 & 5366.82 & 2160.30 & 110.04 \\
\hline

\multirow[c]{3}{*}{\tt Delaunay-N21}
& 5  & 65.32 & 34.66 & 5.13 & 489.07 & 241.18 & 40.69 \\
& 10 & 298.27 & 130.27 & 9.94 & 2350.27 & 1010.31 & 79.50 \\
& 15 & 720.55 & 292.30 & 14.68 & 5719.49 & 2302.70 & 117.80 \\
\hline

\multirow[c]{3}{*}{\tt Epinions}
& 5  & 1.93 & 1.71 & 0.29 & 15.48 & 13.52 & 2.07 \\
& 10 & 4.50 & 3.39 & 0.66 & 35.77 & 26.97 & 4.92 \\
& 15 & 7.06 & 5.05 & 1.03 & 56.26 & 40.65 & 7.93 \\
\hline

\multirow[c]{3}{*}{\tt Facebook}
& 5  & 0.16 & 0.18 & 0.06 & 1.30 & 1.43 & 0.36 \\
& 10 & 0.35 & 0.36 & 0.11 & 2.29 & 2.90 & 0.70 \\
& 15 & 0.54 & 0.54 & 0.15 & 3.63 & 4.36 & 0.92 \\
\hline

\multirow[c]{3}{*}{\tt Gnutella}
& 5  & 1.25 & 1.17 & 0.22 & 9.84 & 9.30 & 1.39 \\
& 10 & 3.24 & 2.33 & 0.51 & 25.77 & 18.56 & 3.56 \\
& 15 & 5.30 & 3.49 & 0.81 & 42.13 & 27.66 & 5.99 \\
\hline

\multirow[c]{3}{*}{\tt HepPh}
& 5  & 0.82 & 0.71 & 0.16 & 6.44 & 5.66 & 1.14 \\
& 10 & 1.93 & 1.42 & 0.37 & 15.37 & 11.31 & 2.72 \\
& 15 & 3.08 & 2.13 & 0.60 & 24.55 & 16.96 & 4.38 \\
\hline

\multirow[c]{3}{*}{\tt HepTh}
& 5  & 0.65 & 0.60 & 0.12 & 5.13 & 4.76 & 0.80 \\
& 10 & 1.50 & 1.19 & 0.27 & 11.97 & 9.52 & 1.86 \\
& 15 & 2.44 & 1.79 & 0.42 & 19.37 & 14.28 & 3.01 \\
\hline

\multirow[c]{3}{*}{\tt Orkut}
& 5  & 251.99 & 244.67 & 75.45 & 1829.83 & 1944.52 & 530.65 \\
& 10 & 679.28 & 484.48 & 170.61 & 5341.31 & 3823.49 & 1294.06 \\
& 15 & 1118.66 & 727.29 & 263.93 & 8701.35 & 5789.93 & 2043.52 \\
\hline

\multirow[c]{3}{*}{\tt Slashdot}
& 5  & 2.04 & 1.70 & 0.32 & 16.14 & 13.52 & 2.28 \\
& 10 & 4.79 & 3.38 & 0.75 & 38.20 & 26.99 & 5.69 \\
& 15 & 7.61 & 5.07 & 1.20 & 60.69 & 40.47 & 9.18 \\
\hline

\multirow[c]{3}{*}{\tt Twitch}
& 5  & 10.29 & 6.52 & 2.50 & 78.06 & 49.68 & 18.28 \\
& 10 & 25.22 & 13.02 & 6.26 & 198.34 & 98.80 & 48.68 \\
& 15 & 40.34 & 18.83 & 10.11 & 320.38 & 145.75 & 78.63 \\
\hline

\cline{1-8}
\multicolumn{2}{l||}{Speedup vs {\tt Naive}}
& $-$ & \rew{$1.45\times$} & \rew{$6.95\times$}
& $-$ & \rew{$1.44\times$} & \rew{$7.41\times$} \\

\multicolumn{2}{l||}{Speedup vs {\tt AVX2}}
& $-$ & $-$ & \rew{$4.79\times$}
& $-$ & $-$ & \rew{$5.15\times$} \\
\end{tabular}
}
\renewcommand{\arraystretch}{1}
\end{table*}

\FloatBarrier

\rew{Table~\ref{tab:rwtime} compares Naive, AVX2, and SABA over the ten datasets and three walk lengths. Across these 30 configurations, AVX2 achieves geometric-mean speedups of $1.45\times$ and $1.44\times$ over Naive for $2048$ and $16{,}384$ walks per vertex, respectively. SABA achieves $6.95\times$ and $7.41\times$ over Naive, and $4.79\times$ and $5.15\times$ over AVX2; its maximum speedup over Naive is approximately $49\times$ in the $16{,}384$-walk setting. These comparisons evaluate streaming sampling, range reduction, and bouquet ordering together. The Baseline/SABA comparison in Section~\ref{sec:aesc-exp} measures the joint benefit within AESC, with shared implementation optimizations included in Baseline.}

\rew{Vectorization advances multiple walks together, streaming selection uses multiply-and-scale range reduction, and bouquet ordering improves memory locality. The synthetic comparisons assess these changes together; the AESC comparison measures their effect within the complete pipeline.}

\subsection{Applicability and Limitations}\label{sec:limits}

\rew{SABA is expected to help when lanes reuse current vertices or nearby adjacency locations.} When duplicates are present in bouquets, the edges can be retrieved from the cache. Otherwise, even with bouquets, the same number of retrievals are done with additional register and memory pressure. \rew{Bouquet ordering is therefore a useful heuristic whose effectiveness depends on the occurrence of duplicates within the processed walks.}
\rew{The bouquet width $B$ is also bounded by the available SIMD lane count; with AVX2 registers, eight 32-bit lanes are processed per instruction.}

For a bouquet with $B$ random walks, let $\beta$ be the number of elements the samples come from and $\eta$ be the number of distinct items in a single level of a bouquet. \rew{To obtain a tractable expression, the lanes are assumed to sample independently and uniformly from a fixed pool of $\beta$ elements; this independent-lane model serves as a diagnostic only, since SABA processes state-dependent walks whose seed-sorted lanes are
conditionally dependent.} Under these assumptions, the ratio of the number of distinct items to the number of selected items can be written as
\rew{\begin{equation}
    \frac{\mathbb{E}[\eta]}{B}
    =
    \frac{\beta}{B}
    \left(
    1-\left(1-\frac{1}{\beta}\right)^B
    \right)
    \approx
    \frac{\beta}{B}
    \left(
    1-e^{-B/\beta}
    \right).
\end{equation}}
Intuitively, independently sampling a small number of vertices from a large number of vertices, e.g. $\beta \gg B$, makes duplicate items not frequent. \rew{Deviations of SABA from these assumptions imply that the formula indicates a trend rather than a bound.}

\rew{Bouquets pay off when the effective support of a level is small or concentrated, so many lanes revisit the same vertices or cache lines; when neighborhoods are large and nearly uniform, lanes rarely converge on the same neighbors and the locality gain disappears. Longer walks spread lane states apart level by level, so the gains concentrate on the early levels.}

\rew{Our experiments cover social, collaboration, and peer-to-peer graphs, together with the California road network and Delaunay-N21 mesh. The two structured networks show some of the largest combined synthetic speedups and substantial cache-miss reductions, so large diameter alone does not imply a small benefit. These synthetic comparisons combine sampling, range reduction, and ordering; they do not isolate ordering. California has a completed AESC benchmark, whereas Delaunay-N21 does not. We therefore distinguish the synthetic results from end-to-end AESC behavior and do not generalize beyond the tested graph families.}
\rew{On almost-regular graphs, nearby initial states can still produce the same or nearby first-step selections. Later steps depend on the visited vertices, state multiplication, and fallback. The independent-lane occupancy model above is a diagnostic rather than a bound on these scheduled walks. The measured synthetic and cache results, rather than degree or diameter alone, characterize the benefit on California and Delaunay-N21.}

\section{Related Work}\label{sec:rel}

In the literature, there exist two ways to improve random walk performance on graphs; first, approximation methods can utilize underlying graph structures to generate diffusion matrices that estimate the probability of random walk visits. {\sc ARK}, proposed by Kang~et~al.~\cite{kernelwalk}, utilizes low-rank structures of graphs and estimates random walk diffusion processes using a few ranks of eigenvectors. In addition, Sarma et al. proposed several algorithms~\cite{Sarma} that employ dynamic programming to improve the asymptotic performance of random-walk problems, including destination-only $k$-random walks, source-destination pair $k$-random walks, and only the path position of vertices from $k$-random walks. In AESC, low-rank eigenvector methods can introduce errors, and they do not perform well for a relatively small number of walks or sources. Existing dynamic programming methods focus on single-source walks and might not be applicable when every step of the walk is required to be processed, i.e., when the vertices are required, in order, as in the case of AESC. \rew{\rew{For spanning centrality, Hayashi et al.~\cite{hayashi2016efficient} sample uniform spanning trees using Wilson\textquotesingle s algorithm, Mavroforakis et al.~\cite{mavroforakis2015spanning} combine random projections with Laplacian solvers, and Peng et al.~\cite{peng2021local} use truncated random walks. {\tt TGT+}~\cite{aesckdd} combines truncated deterministic traversals with random walks while keeping every visited vertex available to the application.} Recently, Liu et al.~\cite{liu2026rwcentrality} proposed near-linear algorithms for random-walk centrality based on incomplete Cholesky factorization and rooted spanning-tree sampling; random-walk centrality is a node hitting-time metric and is not the AESC metric studied here, but it illustrates the continued use of walk sampling in large-scale centrality computation.}

The second category of performance improvements is utilizing compute resources efficiently, {\sc SkyWalker+}, proposed by Wang et al.
\cite{10058015}, uses compressed aliasing and speculative execution to speed up random walk workloads. There also exists a GPU-based method, {\sc C-Saw}, proposed by Pandey~et~al.~\cite{9355289}, that exploits the sampling stage to improve performance and uses inverse transform sampling. A similar approach is applied by {\sc NextDoor}, proposed by Jangda~et~al.~\cite{nextdoor}, via rejection sampling. These methods, even though they may be relevant, do not focus on CPU vectorization and spatial locality. \rew{More recent systems broaden this design space: {\sc LightRW}~\cite{tan2023lightrw} pipelines dynamic graph random walks on FPGAs with degree-aware caching, {\sc FastRW}~\cite{gao2024fastrw} combines dataflow scheduling with hybrid memory on FPGAs, {\sc FlowWalker}~\cite{mei2024flowwalker} employs a memory-efficient GPU sampler with dynamic scheduling, and {\sc GRAPE}~\cite{cappelletti2023grape} provides parallel CPU implementations for graph processing and random-walk-based embedding. These systems focus on high-throughput random-walk generation for graph embedding and machine-learning pipelines, exploiting the parallelism available in GPUs, FPGAs, and modern multicore CPUs.}
\rew{Across both categories, three axes organize this landscape: the underlying approach, either structure-aware approximation or compute-efficient walk execution; the target hardware, ranging from single-threaded CPUs through multicore CPUs, GPUs, and FPGAs; and the graph structures handled well, where approximation methods favor static graphs amenable to eigendecomposition or factorization, \rew{whereas the locality benefit examined here varies across the tested social, collaboration, peer-to-peer, road, and mesh networks (Section~\ref{sec:limits})}.}

\rew{Boundary with our prior work: our influence-maximization kernels~\cite{gokturk2020boosting,GOKTURK2024119875} employ seed-aware, hash-based simulation without mutable per-sample PRNG state and explore the graph in a breadth-first manner. This work extends that principle to random walks: each walk combines its evolving state with the hash of its current vertex, and ordering the walks by these seeds renders their execution locality-focused and depth-first-like, as lanes with similar seeds traverse similar paths through the graph. The contribution of this work is the identification of this bouquet structure together with the locality-focused random-walk processing it enables.}

\section{Conclusion and Future Work}\label{sec:conc}

In this work, we presented a sampling-aware random walk method that clusters paths as {\em bouquets}. We tested the proposed approach on a real-world use case, {\em All Edges Spanning Centrality}. Using the proposed techniques, we provide a fast {\sc AESC} implementation that utilizes multi-threading and {\sc AVX2} instructions. We compared the performance with \rew{the state-of-the-art {\tt TGT+} implementation} on real-world datasets and illustrate that our implementation can be an order of magnitude faster.
\rew{Performance depends on the graph and workload: California and Delaunay-N21 show large gains from the combined sampling and bouquet design, while the California AESC execution scales non-monotonically and Delaunay-N21 has no completed AESC result. The tested graph families do not establish a universal performance guarantee (Section~\ref{sec:limits}).} \rew{Moreover, all results are obtained on Broadwell-era Xeons; wider vectors (AVX-512) and newer core architectures may shift the balance between vectorization and memory-bound behavior.}
\rew{Since GPU execution of Monte-Carlo-based algorithms on graphs using SABA and hash-based sampling raises architecture-specific memory-management issues and would divert attention from the central contribution of this work, namely the bouquet structure, we defer its study to future work.}

\section*{Declarations}

\subsection*{Funding}
No funding was received for this work.

\subsection*{Competing interests}
The authors declare no competing interests.

\subsection*{Data and code availability}
The code, data, and experimental results supporting this study is publicly available at \url{https://github.com/ggokturk/tgtp}.
\subsection*{Author contributions}
Both authors contributed to the review and revision of the manuscript.

\bibliography{refs}

@Inbook{Teixeira2016, author="Teixeira, Andreia Sofia and Santos, Francisco C. and Francisco, Alexandre P.", title="Spanning Edge Betweenness in Practice", bookTitle="Complex Networks VII: Proceedings of the 7th Workshop on Complex Networks CompleNet 2016", year="2016", publisher="Springer International Publishing", address="Cham", pages="3--10", isbn="978-3-319-30569-1", doi="10.1007/978-3-319-30569-1_1", url="https://doi.org/10.1007/978-3-319-30569-1_1" }

@article{GOKTURK2024119875,
title = {Fast and error-adaptive influence maximization based on Count-Distinct sketches},
journal = {Information Sciences},
volume = {655},
pages = {119875},
year = {2024},
issn = {0020-0255},
doi = {https://doi.org/10.1016/j.ins.2023.119875},
url = {https://www.sciencedirect.com/science/article/pii/S0020025523014603},
author={G{\"o}kt{\"u}rk, G{\"o}khan and Kaya, Kamer}
}

@article{pearson1905problem,
  title={The problem of the random walk},
  author={Pearson, Karl},
  journal={Nature},
  volume={72},
  number={1865},
  pages={294--294},
  year={1905},
  publisher={Nature Publishing Group UK London}
}

@article{gokturk2020boosting,
  title={Boosting parallel influence-maximization kernels for undirected networks with fusing and vectorization},
  author={G{\"o}kt{\"u}rk, G{\"o}khan and Kaya, Kamer},
  journal={IEEE Transactions on Parallel and Distributed Systems},
  volume={32},
  number={5},
  pages={1001--1013},
  year={2020},
  publisher={IEEE}
}

@inproceedings{aesckdd, author = {Zhang, Shiqi and Yang, Renchi and Tang, Jing and Xiao, Xiaokui and Tang, Bo}, title = {Efficient Approximation Algorithms for Spanning Centrality}, year = {2023}, isbn = {9798400701030}, publisher = {Association for Computing Machinery}, address = {New York, NY, USA}, url = {https://doi.org/10.1145/3580305.3599323}, doi = {10.1145/3580305.3599323}, booktitle = {Proceedings of the 29th ACM SIGKDD Conference on Knowledge Discovery and Data Mining}, pages = {3386–3395}, numpages = {10}, location = {, Long Beach, CA, USA, }, series = {KDD '23} }

@inproceedings{peng2021local,
  title={Local algorithms for estimating effective resistance},
  author={Peng, Pan and Lopatta, Daniel and Yoshida, Yuichi and Goranci, Gramoz},
  booktitle={Proceedings of the 27th ACM SIGKDD Conference on Knowledge Discovery \& Data Mining},
  pages={1329--1338},
  year={2021}
}

@inproceedings{hayashi2016efficient,
  title={Efficient Algorithms for Spanning Tree Centrality.},
  author={Hayashi, Takanori and Akiba, Takuya and Yoshida, Yuichi},
  booktitle={IJCAI},
  volume={16},
  pages={3733--3739},
  year={2016}
}

@inproceedings{mavroforakis2015spanning,
  title={Spanning edge centrality: Large-scale computation and applications},
  author={Mavroforakis, Charalampos and Garcia-Lebron, Richard and Koutis, Ioannis and Terzi, Evimaria},
  booktitle={Proceedings of the 24th international conference on world wide web},
  pages={732--742},
  year={2015}
}

@inproceedings{teixeira2013spanning,
  title={Spanning edge betweenness},
  author={Teixeira, Andreia Sofia and Monteiro, Pedro T and Carri{\c{c}}o, Jo{\~a}o A and Ramirez, M{\'a}rio and Francisco, Alexandre P},
  booktitle={Workshop on mining and learning with graphs},
  volume={24},
  pages={27--31},
  year={2013},
  organization={Citeseer}
}

@misc{snapnets,
author       = {Jure Leskovec and Andrej Krevl},
title        = {{SNAP Datasets}: {Stanford} Large Network Dataset Collection},
howpublished = {\url{http://snap.stanford.edu/data}},
month        = jun,
year         = 2014
}

@ARTICLE{10058015,
  author={Wang, Pengyu and Xu, Cheng and Li, Chao and Wang, Jing and Wang, Taolei and Zhang, Lu and Hou, Xiaofeng and Guo, Minyi},
  journal={IEEE Transactions on Computers}, 
  title={Optimizing {GPU}-Based Graph Sampling and Random Walk for Efficiency and Scalability}, 
  year={2023},
  volume={72},
  number={9},
  pages={2508-2521},
  doi={10.1109/TC.2023.3251860}}

@inproceedings{kernelwalk,
  title={Fast random walk graph kernel},
  author={Kang, U and Tong, Hanghang and Sun, Jimeng},
  booktitle={Proceedings of the 2012 SIAM international conference on data mining},
  pages={828--838},
  year={2012},
  organization={SIAM}
}

@article{Sarma,
author = {Das Sarma, Atish and Nanongkai, Danupon and Pandurangan, Gopal and Tetali, Prasad},
title = {Distributed Random Walks},
year = {2013},
issue_date = {February 2013},
publisher = {Association for Computing Machinery},
address = {New York, NY, USA},
volume = {60},
number = {1},
issn = {0004-5411},
url = {https://doi.org/10.1145/2432622.2432624},
doi = {10.1145/2432622.2432624},
journal = {J. ACM},
month = {feb},
articleno = {2},
numpages = {31}
}

@inproceedings {9355289,
author = {S. Pandey and L. Li and A. Hoisie and X. Li and H. Liu},
booktitle = {2020 SC20: International Conference for High Performance Computing, Networking, Storage and Analysis (SC)},
title = {C-SAW: A Framework for Graph Sampling and Random Walk on {GPU}s},
year = {2020},
volume = {},
issn = {},
pages = {780-794},
doi = {10.1109/SC41405.2020.00060},
url = {https://doi.ieeecomputersociety.org/10.1109/SC41405.2020.00060},
publisher = {IEEE Computer Society},
address = {Los Alamitos, CA, USA},
month = {nov}
}

@article{nextdoor,
  title={NextDoor: {GPU}-Based Graph Sampling for Graph Machine Learning},
  author={Abhinav Jangda and Sandeep Polisetty and Arjun Guha and Marco Serafini},
  journal={ArXiv},
  year={2020},
  volume={abs/2009.06693},
  url={https://api.semanticscholar.org/CorpusID:221703383}
}

@article{matsumoto1998mersenne,
  title={Mersenne twister: a 623-dimensionally equidistributed uniform pseudo-random number generator},
  author={Matsumoto, Makoto and Nishimura, Takuji},
  journal={ACM Transactions on Modeling and Computer Simulation (TOMACS)},
  volume={8},
  number={1},
  pages={3--30},
  year={1998},
  publisher={ACM New York, NY, USA}
}

@misc{murmur3,
  author = {Appleby, Austin},
  title = {{MurmurHash3} reference implementation in {SMHasher}},
  howpublished = {\url{https://github.com/aappleby/smhasher/blob/master/src/MurmurHash3.cpp}},
  year = 2016
}

@article{cappelletti2023grape,
  author={Cappelletti, Luca and Fontana, Tommaso and Casiraghi, Elena and Ravanmehr, Vida and Callahan, Tiffany J. and Cano, Carlos and Joachimiak, Marcin P. and Mungall, Christopher J. and Robinson, Peter N. and Reese, Justin and Valentini, Giorgio},
  title={{GRAPE} for fast and scalable graph processing and random-walk-based embedding},
  journal={Nature Computational Science},
  volume={3},
  pages={552--568},
  year={2023}
}

@article{tan2023lightrw,
  author={Tan, Hongshi and Chen, Xinyu and Chen, Yao and He, Bingsheng and Wong, Weng-Fai},
  title={{LightRW}: {FPGA} accelerated graph dynamic random walks},
  journal={Proceedings of the ACM on Management of Data},
  volume={1},
  number={1},
  pages={90:1--90:27},
  year={2023}
}

@article{mei2024flowwalker,
  author={Mei, Junyi and Sun, Shixuan and Li, Chao and Xu, Cheng and Chen, Cheng and Liu, Yibo and Wang, Jing and Zhao, Cheng and Hou, Xiaofeng and Guo, Minyi and He, Bingsheng and Cong, Xiaoliang},
  title={{FlowWalker}: A memory-efficient and high-performance {GPU}-based dynamic graph random walk framework},
  journal={Proceedings of the VLDB Endowment},
  volume={17},
  number={8},
  pages={1788--1801},
  year={2024}
}

@article{gao2024fastrw,
  author={Gao, Yingxue and Wang, Teng and Gong, Lei and Wang, Chao and Hu, Yiqing and Yang, Yi and Liu, Zhongming and Li, Xi and Zhou, Xuehai},
  title={Enhancing graph random walk acceleration via efficient dataflow and hybrid memory architecture},
  journal={IEEE Transactions on Computers},
  volume={73},
  number={3},
  pages={887--901},
  year={2024}
}

@article{liu2026rwcentrality,
  author={Liu, Changan and Xie, Zixuan and Zehmakan, Ahad N. and Zhang, Zhongzhi},
  title={Efficient algorithms for computing random walk centrality},
  journal={IEEE Transactions on Knowledge and Data Engineering},
  volume={38},
  number={1},
  pages={235--247},
  year={2026}
}

\end{document}